# Definitive upper bound on the negligible contribution of quasars to cosmic reionization

Linhua Jiang[1,2], Yuanhang Ning[1,2], Xiaohui Fan[3], Luis C. Ho[1,2], Bin Luo[4], Feige Wang[3], Jin Wu[1,2], Xue-Bing Wu[1,2], Jinyi Yang[3], Zhen-Ya Zheng[5]

[1]Kavli Institute for Astronomy and Astrophysics, Peking University, Beijing, China
[2]Department of Astronomy, School of Physics, Peking University, Beijing, China
[3]Steward Observatory, University of Arizona, Tucson, Arizona, USA
[4]School of Astronomy and Space Science, Nanjing University, Nanjing, China
[5]Shanghai Astronomical Observatory, Shanghai, China

**Cosmic (hydrogen) reionization marks one of the major phase transitions of the universe at redshift $z \geq 6$. During this epoch, hydrogen atoms in the intergalactic medium (IGM) were ionized by Lyman continuum (LyC) photons. However, it remains challenging to identify the major sources of the LyC photons responsible for reionization. In particular, individual contributions of quasars (or active galactic nuclei, AGNs) and galaxies are still under debate. Here we construct the far-ultraviolet (far-UV) luminosity function for type 1 quasars at $z \geq 6$ that spans 10 magnitudes ($-19 \leq M_{UV} \leq -29$), conclusively showing that quasars made a negligible contribution to reionization. We mainly search for quasars in the low-luminosity range of $M_{UV} > -23$ mag that is critical to determine quasars' total LyC photon production but has been barely explored previously. We find that the quasar population can only provide less than 7% (95% confidence level) of the total photons needed to keep the universe ionized at $z = 6.0 - 6.6$. Our result suggests that galaxies, presumably low-luminosity star-forming systems, are the major sources of hydrogen reionization.**

High-redshift ($z \geq 6$) galaxies and quasars (here referring to broad-line AGNs regardless of luminosity) are thought to be two major candidate sources of LyC photons at the epoch of cosmic reionization[1-4] (EoR). Their individual contributions to reionization are unclear[5-9]. The current understanding of galaxies' contribution to reionization is mainly hampered by our limited knowledge about the escape fraction ($f_{esc}$) of LyC photons from galaxies to the IGM. Observations of LyC at $z \gtrsim 3$ have yielded non-detections for most galaxies observed so far, and the average $f_{esc}$ is only a few percent[10,11]. A higher $f_{esc}$ is needed for reionization, depending on the complex, intrinsic LyC production process that is poorly known[3,12,13]. Due to the IGM absorption, direct LyC detection from high-redshift galaxies is unrealistic. For quasars, little is known about their faint population at $z \geq 6$ that may play an important role. Type 1 quasar luminosity function (QLF) brighter than $M_{1450} \approx -23$ mag (absolute magnitude at rest-frame 1450 Å) has been established[14-18], and a simple extrapolation of this QLF to $M_{1450} \approx -18$ mag has suggested a small contribution from quasars. However, studies of X-ray-detected objects inferred a much higher spatial density of faint quasar candidates at $3 \leq z \leq 6$ than previous measurements, leading to a claim that quasars likely provided a substantial fraction of ionizing photons required by reionization[6,7]. As the debate is on-going[19-21], the key is to constrain the faint-end of the QLF in the far-UV range that is directly connected to the overall quasar LyC production. In this paper, we assume $H_0 = 70$ km s$^{-1}$ Mpc$^{-1}$, $\Omega_m = 0.3$, and $\Omega_\Lambda = 0.7$, where $H_0$ is the current value of the Hubble constant, and $\Omega_m$ and $\Omega_\Lambda$ are the cosmological density parameters for matter and dark energy, respectively.

We searched for low-luminosity type 1 quasars at $6.0 \leq z \leq 6.6$ using a combination of ground-based and Hubble Space Telescope (HST) images. Table 1 and Extended Data Table 1 list the fields and data sets that we used. The specific range of $6.0 \leq z \leq 6.6$ is optimal for our purpose. First of all, the bright-end of the QLF in this range has been well determined, and the quasar density during EoR reaches its maximum in this range. In addition, these quasars have substantial flux in the observed-frame optical so that HST Advanced Camera for Surveys (ACS) images can be used to efficiently separate point-like quasars from extended objects. Finally, quasars in this redshift range can be easily distinguished from their major point-source contaminants, Galactic L/T dwarf stars. Fig. 1a compares the spectra of a $z = 6.2$ quasar, a L6 star, and a T6 star (see details in Methods). The quasar spectrum appears very different from those of the stars in the wavelength range covered by the $i$, $z$, and $J$-band filters ("$izJ$" denote a few slightly different filter sets in Table 1). In particular, the Lyman forest absorption in the quasar spectrum generates a strong Ly$\alpha$ break that produces a very red $i-z$ color (magnitudes are on the AB system). Meanwhile, the $z-J$ color of the quasar is much bluer than those of the stars. Therefore, type 1 quasars at $6.0 \leq z \leq 6.6$ are well separated from L/T dwarfs in the $z-J$ versus $i-z$ color-color diagram (Fig. 1b; see also Extended Data Fig. 1). We caution that both "redshift" and "$z$-band" are expressed as "$z$" in the paper, but the meaning of "$z$" should be clear from the context in situ.

Our quasar selection procedure consists of two major steps, the color selection mentioned above and a morphological selection illustrated in Fig. 2 (see also Methods). We aim to detect type 1 quasars whose rest-frame far-UV radiation is dominated by central AGNs. Due to the $(1+\text{redshift})^4$ surface brightness dimming, these quasars would appear as point-like sources in HST ACS images. We used the compactness of the sources, as judged by their full-width at half-maximum (FWHM), to separate quasars from their major extended-source contaminants, galaxies with similar $i-z$ and $z-J$ colors at similar redshifts. Fig. 2 demonstrates this selection in the GOODS fields. The grey dots represent sources brighter than $z = 27$ mag (10 $\sigma$ detection limit), and point sources occupy a distinct locus at FWHM $\approx 0.1$ arcsecs (''). To select quasars brighter than $z = 26$ mag, we added a tolerance of 0.03'' (1 pixel size) so that our FWHM cut is at 0.13''. The point-source locus is blurred toward the faint end due to the increasing measurement uncertainties. We included this effect using simulations, and the result is shown as the pink curve at $z = 25.8$-27.0 mag. About 95% of the simulated point sources are below this curve. We adopt this curve as our selection criterion for the range of 26-27 mag. The blue circles in Fig. 2 represent $i$-band dropout objects with $i - z > 2$ mag. None of them was detected in the deepest, 7 Ms Chandra X-ray image[36]. Four of them satisfy our FWHM criterion (i.e., below the FWHM selection line), but their colors are consistent with those of L/T dwarfs. We did not find high-redshift quasars in the two fields down to the limit of $z = 27$ mag.

Cosmological simulations have shown that the host galaxies of high-redshift quasars are very compact[37,38]. Despite this fact, the usage of FWHM and the 0.03'' tolerance ensures that we are able to select quasars even if they have some extended radiation from host galaxies in the far-UV. We performed a series of simulations using HST ACS images. We first made a point-spread function (PSF) by stacking bright point sources in an ACS image. We then generated a set of model galaxies with difference sizes and Sérsic profiles (Methods). A model quasar was the combination of the PSF and a model galaxy. By scaling the relative and absolute flux of the two components, we built a large sample of mock quasars. We randomly inserted the mock quasars to the ACS image and measured their FWHMs. The dashed line in Fig. 2 shows one result, in which host galaxies contribute 30% of the total far-UV radiation (30% is more than 5-10 times higher than that found in bright $z \sim 6$ quasars[39]). This line is well below our FWHM selection line. If host galaxies are star-forming galaxies, they have little effect on our color selection because they have similar $izJ$ colors as those of quasars. If they are not star-forming galaxies, they emit little far-UV radiation. On the other hand, the UV radiation from a quasar and its host is indistinguishable, so we assume that all UV radiation is from the quasar and neglect the host component in the following analyses. Consequently, the quasar contribution to reionization will be slightly overestimated. In Fig. 2, we also plot a sample of spectroscopically confirmed galaxies at $z > 5.9$ (refs[33-35]). One of them is located at the FWHM selection boundary, and the remaining galaxies are all above the selection line, indicating that our selection suffers little contamination from galaxies.

Our search for quasars in the imaging fields listed in Table 1 did not yield any candidate at $z \geq 6$ (Methods). This result is broadly consistent with previous observations of high-redshift galaxies. In the past two decades, a number of deep spectroscopic observations were conducted to identify Lyman-break galaxies (LBGs) at $6 \leq z \leq 7$ in these fields, and the deepest observations reached $z > 27$ mag for Lyman-$\alpha$ emitters (LAEs) (refs[33-35,40,41]). In these observations, LBG targets were usually $i$-band dropouts without other color or morphological constraints, meaning that they potentially included all quasar candidates of our type at similar redshifts. No quasars have been reported to date. While it is difficult to characterize the depth and completeness of these observations for high-redshift quasars, their results support our conclusion that there are few type 1 quasars in these fields down to the limits that we have probed. We further used our own spectroscopic programs[40,41] (no. 7 and no. 8 in Table 1) to search for faint quasars. In these two programs, the LBG targets were also $i$-band dropouts, including many objects found in the above imaging fields. We carried out deep spectroscopic observations of the $i$-band dropouts using Keck and Magellan telescopes, and we did not find high-redshift quasars (Methods).

Based on the above imaging and spectroscopic results, we measure quasar spatial densities and derive the QLF at $6.0 \leq z \leq 6.6$. For each field, we first use simulations to calculate its sample completeness or selection function, defined as the probability that a quasar with a given magnitude $M_{1450}$ and redshift $z$ can be selected. The procedure is the same as those used in the literature[15,16]. Briefly, we generate a grid of model quasars whose spectral energy distributions (SEDs) match the observed SEDs of quasars, and then calculate the fraction of quasars that satisfy our color-selection criteria. Extended Data Fig. 2 shows four representative selection functions. We then compute the spatial density using the $1/V_a$ method. The available (or accessible) volume $V_a$ for a quasar with $M_{1450}$ and $z$ in a magnitude bin $\Delta M$ and a redshift bin $\Delta z$ is $\int_{\Delta M} \int_{\Delta z} p(M_{1450}, z) \frac{dV}{dz} dz \, dM$, where $p(M_{1450}, z)$ is the selection function. The differential density at redshift $z$ is $\phi(M_{1450}) = \sum_i 1/V_a^i$, where the sum is over all quasars. Since we did not find any quasars, we adopt the cumulative density $\phi(<M_{1450})$ that is the integral of the differential density over the whole magnitude range covered by the field. In this case, the available volume is the integral of the above $V_a$ over the magnitude range, and the numbers of quasars are upper limits. We will use two confidence levels (CLs) for upper limits, 75% and 95%. The upper limits of a zero detection at the two CLs are 1.386 and 2.996, respectively[42], meaning that we actually use 1.386 and 2.996 as the numbers of quasars in our calculations. As seen from Extended Data Fig. 2, a fixed flux limit corresponds to slightly different $M_{1450}$ limits at different redshifts. We define an effective magnitude limit for each field. This limit is determined so that the total probability integrated over the $M_{1450}$ range in the selection function is the same as the total probability brighter than this limit with a 100% completeness. The definition of this limit does not affect the available volume.

The results of the cumulative densities for the individual fields are shown as the open circles in Fig. 3a. They are the upper limits for non-detections at the CL of 75%. Most individual fields do not put stringent constraints on the quasar density due to their small area. To improve the constraints, we divide all fields into four groups based on their survey depths (Methods). Compared to the individual fields, these groups have the same upper limits (non-detections), but with larger survey volumes. The results are illustrated as the filled circles and squares, showing four upper limits at the CLs of 75% and 95%, respectively. The combined measurements provide strong constraints on the faint end of the QLF. The triangles in Fig. 3a represent the cumulative densities of bright quasars computed from their differential densities in a previous study[16] (Methods). As seen from the figure, the bright end of the QLF is well described by the five brightest data points. The combination of these five data points and our results serves as two fiducial, observed QLFs (CL = 75% and 95% at the faint end). To minimize the impact of the previous study on the measurement of the faint-end QLF, we do not include the two faintest data points (open triangles) from the previous study. Fig. 3a shows that we are able to directly constrain the far-UV QLF at $z > 6$ down to $M_{1450} \approx -19$ mag.

We model the differential QLF using a double power-law (DPL) that has four parameters, including the faint-end slope $\alpha$, bright-end slope $\beta$, characteristic magnitude $M^*$, and normalization factor $\phi^*$. The bright-end slope $\beta$ is fixed to be $-2.73$, as it has been well determined[15,16]. To find the best model fits for the two free parameters $\alpha$ and $M^*$, we use a grid search and build a large grid of $[\alpha, M^*]$ values. For each $[\alpha, M^*]$ pair, we calculate cumulative densities from the differential model QLF, and compare them with our observed, fiducial QLFs by minimizing $\chi^2$. We then locate the minimum of all minimized $\chi^2$ values to search for the best fit of $[\alpha, M^*]$ and its associated $\phi^*$. The two solid curves in Fig. 3a are the best model fits to the two fiducial QLFs. There is no previous, direct measurement of the far-UV QLF at $z > 6$ that covers the similar luminosity range, so we compare our work with previous results from quasar evolution models. These models are usually calibrated against observational measurements. The two green curves in Fig. 3a represent two previous results[43,44]. They are broadly consistent with our results.

We use a standard method[15,16] to calculate the number of ionizing photons provided by quasars and estimate the quasar contribution to the ionizing UV background at $z > 6$. The calculation is done with the two best-fit models for the two fiducial QLFs. We assume an $f_{esc}$ of 75% (ref[45]) and a broken power-law quasar SED[46] around 912 Å: $f_\nu \propto \nu^{-0.6}$ at wavelength $\lambda > 912$ Å and $f_\nu \propto \nu^{-1.7}$ at $\lambda < 912$ Å, where $\nu$ is frequency. We then integrate the SED over an energy range of 1–4 Ryd, integrate the QLF over a luminosity range from $M_{1450} = -30$ to $-18$ mag, and compute the total photon emissivity required to ionize the universe at $z \approx 6.2$. An integration to a lower magnitude limit has little effect on the result because of the flat slope at the faint end. We have also assumed that the quasar UV radiation is isotropic, so the contribution of type 2, obscured quasars is considered. The total photon emissivity per unit comoving volume required to ionize the universe is $\dot{N}_{ion}(z) = 10^{50.48}(\frac{C}{3}) \times (\frac{1+z}{7})^3$ Mpc$^{-3}$ s$^{-1}$ (ref[47]), where $C$ is the IGM clumping factor, for which we adopt a typical value 3 (ref[48]), and we have assumed that the baryon density $\Omega_b h^2$ is 0.022. The solid curves in Fig. 3b illustrate the calculated fractions ($f_{AGN}$) of the quasar emissivity to the required photon emissivity. The results for the two cases are ~3 % and 7%, respectively. They give a definitive upper bound of $f_{AGN} \approx 7\%$ at the CL of 95%.

We have provided direct evidence for a negligible quasar contribution to reionization. The quasar contribution depends on a few parameters. For example, if we decrease $C$ to 2 or increase $f_{esc}$ to 100%, the upper bound slightly increases to $f_{AGN} \approx 10\%$. There is evidence that faint quasars may have much lower $f_{esc}$ than what we adopted earlier. If we assume $f_{esc} = 30\%$ (ref[49]), the upper bound of $f_{AGN}$ decreases to ~3%. In addition, we have used a fixed SED around 912 Å. It has been found that SEDs from evolving thin accretion disks can produce up to ~80% more ionizing photons[50]. This would increase $f_{AGN}$ by up to ~80% and the upper bound would change to ~12% (in case of $f_{esc} = 75\%$). In all these cases, the quasar contribution to reionization remains small. In the near future, the combination of deep optical images from the China Space Station Telescope[51] and near-IR images from the Euclid and Roman Space Telescopes will provide a stronger constraint on $f_{AGN}$.

Our conclusion suggests that galaxies (stars) are the only major candidate of the reionization sources if we neglect non-standard candidates. Deep HST observations have implied that faint star-forming galaxies can potentially provide sufficient ionizing photons if their $f_{esc}$ is higher than those in low-redshift galaxies. This claim was overshadowed by recent studies of high-redshift quasar spectra that show a short mean free path of ionizing photons at $z \approx 6$ (ref[52]), which demands an increase of ionizing photons to keep the universe ionized[53]. If this is confirmed, our result further implies that galaxies at $z \geq 6$, compared to their low-redshift counterparts, have higher $f_{esc}$ and/or a higher production rate of ionizing photons from low-metallicity stellar populations.

**Data availability** All imaging data used in this paper are publicly available and the details are presented in Table 1 and Methods.

**Code availability** Data were reduced using publicly available data reduction pipelines.

**Acknowledgements** L.J., Y.N., L.H., J.W., and X.W. acknowledge support from the National Science Foundation of China (11721303, 11890693, 12022303) and the China Manned Space Project (CMS-CSST-2021-A04, CMS-CSST-2021-A05, CMS-CSST-2021-A06). This paper used observations made with the NASA/ESA Hubble Space Telescope, and obtained from the Hubble Legacy Archive, which is a collaboration between the Space Telescope Science Institute (STScI/NASA), the Space Telescope European Coordinating Facility (ST-ECF/ESA) and the Canadian Astronomy Data Centre (CADC/NRC/CSA).

**Author contributions** L.J. designed the program, analyzed the data, and prepared the manuscript. Y.N. performed the image simulations. L.H. helped to prepare the manuscript. All authors helped with the scientific interpretations and commented on the manuscript.

**Competing interests** The authors declare that they have no competing interests.

**Table 1. Data used for the quasar searches.**

| No. | Field | Filters (izJ) used for color selection | Filter used to measure FWHM | Redshift range | Magnitude limit (mag) | Area (arcmin$^2$) |
|---|---|---|---|---|---|---|
| 1a | GOODS-S and GOODS-N[22] | HST ACS F775W and F850LP, HST WFC3 F125W | ACS F850LP | 6.0 - 6.7 | z = 27.0 | 303 |
| 1b | GOODS-S Deep | HST ACS F775W and F850LP, HST WFC3 F125W | ACS F850LP | 6.0 - 6.7 | z = 27.4 | 23 |
| 2a | COSMOS[23] | Subaru Hyper Suprime-Cam (HSC[24]) i and z, VLT UltraVISTA[25] near-IR | ACS F814W | 6.0 - 6.5 | z = 25.4; F814W ≈ 26.7 | 4977 |
| 2b | COSMOS CANDELS[26] | Subaru HSC i and z, HST WFC3 F125W or VLT UltraVISTA near-IR | ACS F814W | 6.0 - 6.5 | z = 26.1; F814W ≈ 27.4 | 179 |
| 2c | COSMOS Shallow | Subaru HSC i and z, VLT UltraVISTA near-IR | ACS F814W | 6.0 - 6.5 | z = 25.0; F814W ≈ 26.3 | 945 |
| 3a | EGS[27] | CFHT Legacy Survey (CFHTLS) i and z; no near-IR data needed | ACS F814W | 6.0 - 6.5 | z = 25.5; F814W ≈ 26.8 | 342 |
| 3b | EGS CANDELS | CFHTLS i, HST ACS F814W and WFC3 F125W | ACS F814W | 6.0 - 6.5 | z ≈ 26.2; F814W ≈ 27.5 | 205 |
| 4 | UDS CANDELS | Subaru HSC i and z, HST WFC3 F125W and/or UKIDSS UDS[28] near-IR | ACS F814W | 6.0 - 6.5 | z = 26.1; F814W ≈ 27.4 | 222 |
| 5 | GEMS[29] | Subaru HSC i, HST ACS F850LP, VLT VIDEO[30] near-IR | ACS F850LP | 6.0 - 6.7 | z = 26.0 | 660 |
| 6 | Other HST fields | See details in Extended Data Table 1 | | | | |
| 7 | Subaru Deep Field (SDF) | Subaru SCam i and z; J-band data not used | None (spectroscopy) | 6.0 - 6.6 | z ≈ 25.0 | 340 |
| 8 | Magellan M2FS survey | Subaru SCam i and z; J-band data not used | None (spectroscopy) | 6.0 - 6.6 | z ≈ 25.0 | 4138 |

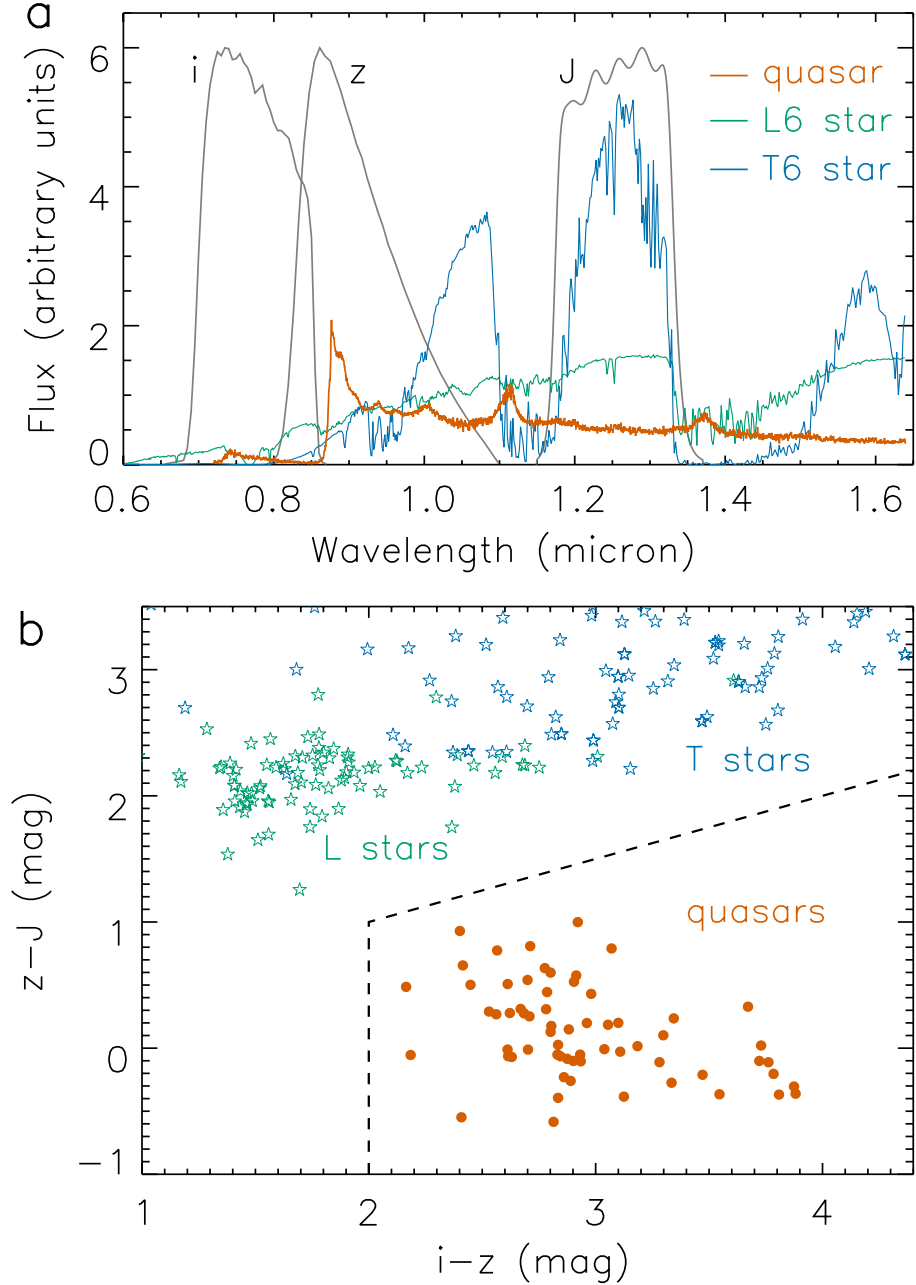

**Fig. 1 Color selection of high-redshift quasars.** (a) Spectra of a quasar at $z = 6.2$ (orange), a L6 star (green), and a T6 star (blue). The three filter ($i$, $z$ and $J$) transmission curves in grey represent the HST ACS F775W and F850LP filters and the UKIRT WFCAM $J$-band filter, respectively. They have been normalized so that the highest values are 6. The quasar spectrum is constructed from two composite spectra[17,31], and the spectra of stars are two model spectra[32] (Methods). The three spectra have been scaled so that they have the same $z$-band magnitudes. (b) The $z−J$ versus $i−z$ color-color diagram. The green and blue stars represent a sample of known L/T dwarfs that cover a wide range of sub-types. The orange dots represent a sample of quasars at redshift between 5.9 and 6.5 (ref[17]). Quasars in this redshift range have distinct colors and thus can be well separated from L/T dwarfs using the dashed lines.

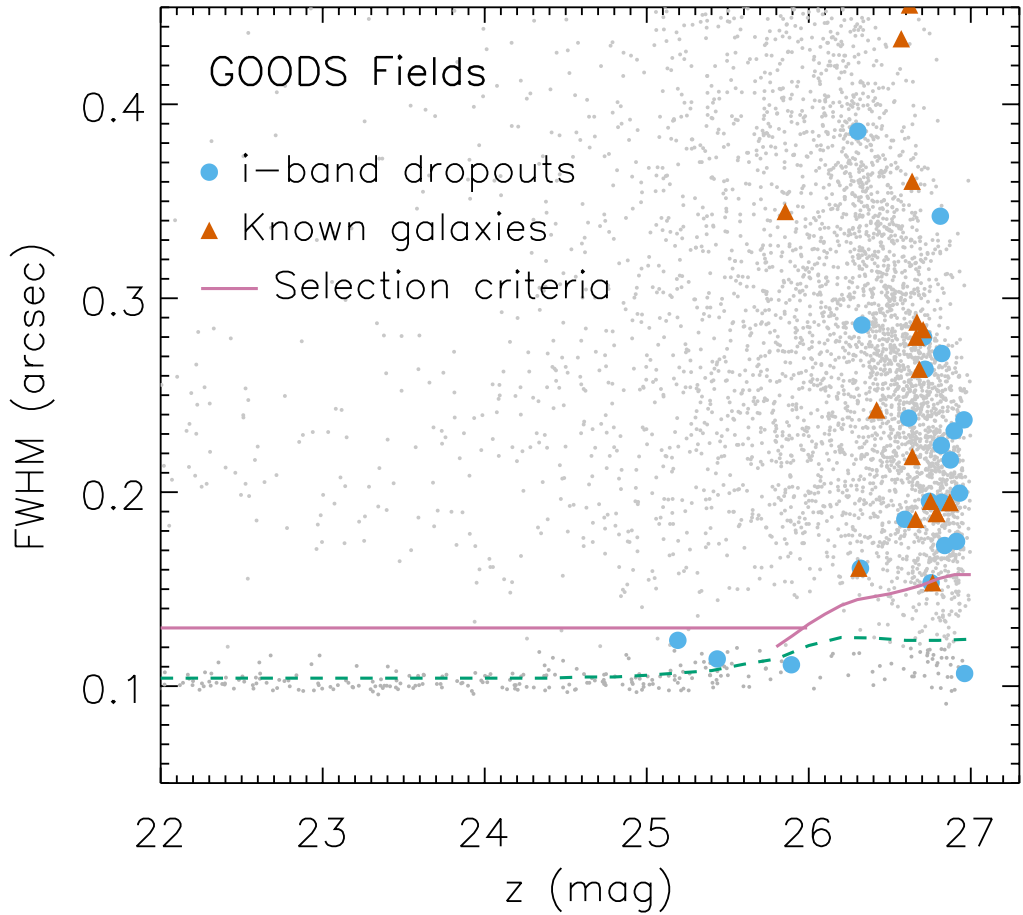

**Fig. 2 Morphological selection of high-redshift quasars.** The grey dots represent objects brighter than $z = 27$ mag ($\sim 10\ \sigma$ detection) detected in the GOODS fields. The point-source locus at FWHM $\approx 0.1''$ is distinct. The pink straight line at $z < 26$ mag and the pink curve at $z > 26$ mag are our FWHM selection criteria. The blue circles represent the *i*-band dropouts selected in the two GOODS fields. The four *i*-band dropouts below the selection line have colors consistent with those of L/T dwarfs. Another one at the selection boundary is a known galaxy at $z \approx 6$ (ref[33]). The red triangles represent a sample of spectroscopically confirmed galaxies at $z > 5.9$ in the GOODS, UDS, and COSMOS fields[33-35]. They are all above the selection line (except the one at the boundary), indicating that our selection suffers little contamination from galaxies. The dashed line is the median track of the FWHM values for a sample of simulated quasars in which host galaxies contribute 30% of the total far-UV radiation. It is well below our FWHM selection line (see Methods for more details).

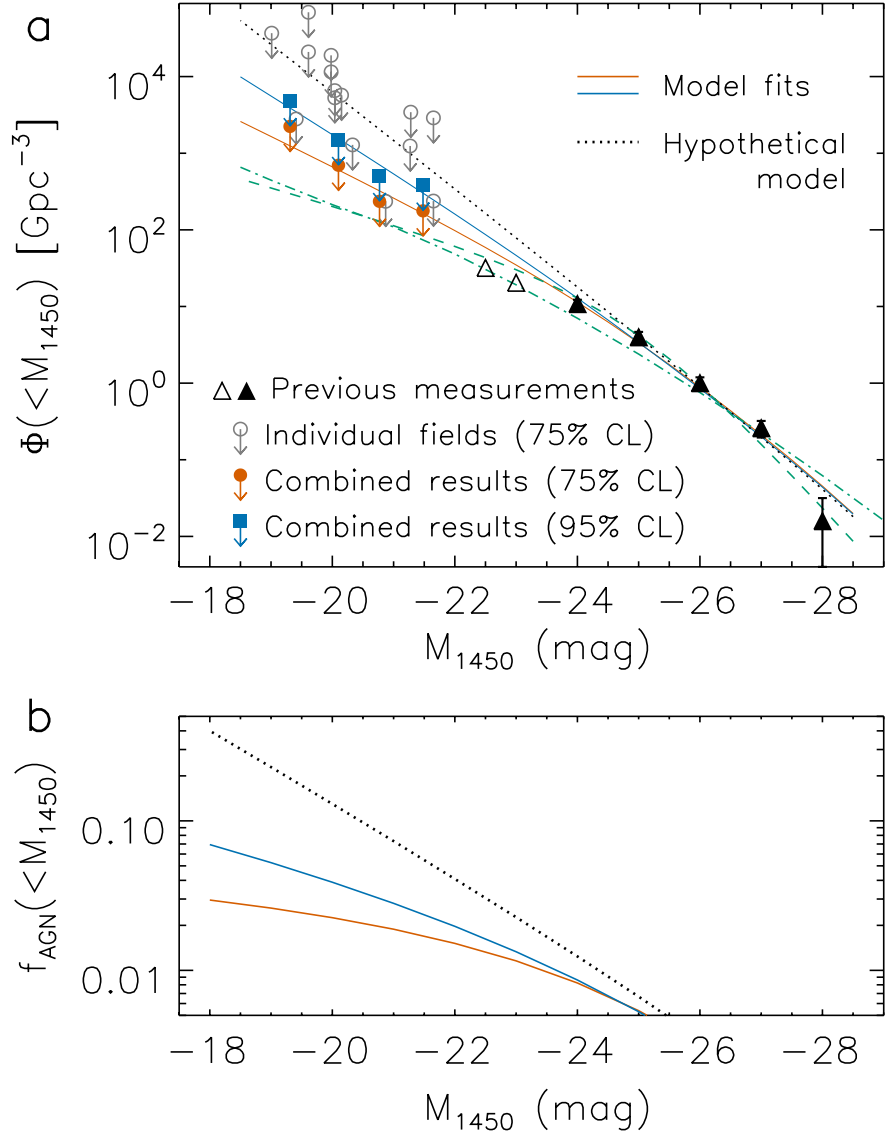

**Fig. 3 QLF and quasar contribution to reionization**. (a) Cumulative QLF at $z \sim 6.2$. The open circles represent the upper limits of our measurements at the CL of 75% for the individual fields. The filled circles and squares represent the upper limits of the four combined measurements at the CLs of 75% and 95%, respectively. They have been shifted by 0.1 mag for clarity. The solid curves are the best fits of a DPL model to the two fiducial QLFs. The open and filled triangles represent previous measurements of bright quasars[16], and the open triangles were not used for the model fitting. The green dashed and dash-dotted curves represent two previous results from quasar evolution models[43,44]. The black dotted line is a hypothetical DPL model in which the faint-end slope is tweaked so that the total quasar contribution reaches ~40% (dotted line in the lower panel). This hypothetical model demonstrates the importance of faint quasars for the calculation of $f_{AGN}$. (b) Quasar contribution to reionization. The three curves represent the fractions ($f_{AGN}$) of the cumulative quasar emissivity to the total photon emissivity required to ionize the universe. They are computed using the best-fit models in the upper panel. They indicate a negligible quasar contribution.

## Methods

### Quasar selection method

Our quasar selection procedure consists of two major steps, a color selection in the $z−J$ versus $i−z$ color-color diagram as shown in Fig. 1 and a morphological selection in the FWHM versus $z$-magnitude diagram as shown in Fig. 2. One may switch the two steps. In Fig. 1a, the quasar spectrum is the combination of two composite spectra, including an optical spectrum[17] at wavelength $\lambda < 1.05$ $\mu$m and a near-IR spectrum[31] at $\lambda > 1.05$ $\mu$m. The spectra of the L6 and T6 dwarfs are model spectra[32], and the specific model parameters are solar metallicity, $\log(g) = 5.0$ (gravity g in units of cm s$^{-2}$), and effective temperature $T_{eff} = 1600$ and 900 K (for L6 and T6, respectively). The model spectra are for the purpose of demonstration only, so the details of the parameters are not important here. In Fig. 1b, the known L/T dwarf stars are from an online archive (DwarfArchives.org; https://www.ipac.caltech.edu/project/dwarf-archives). They cover a wide range of sub-types from L1 to T9. The known quasars at $5.9 < z < 6.5$ are detected by the Pan-STARRS1 survey[17]. We have converted the $i$- and $z$-band magnitudes from the Pan-STARRS1 system to the HST ACS system (see Table 1). We have excluded those without available $J$-band photometry. We have also updated the photometry of a few faint quasars with better photometric measurements[14,54,55].

The optimal redshift range that we chose depends on the combination of the $izJ$ filters. In particular, we ensured that the Ly$\alpha$ emission line has almost completely moved out of the $i$-band filter so that the $i−z$ color is very red. Since the different data sets (or deep fields) that we used have different combinations of the three filters (see Table 1), we derived optimal redshift ranges for individual data sets. The results are demonstrated in Extended Data Fig. 1. In this figure, the color-coded lines show the median tracks of the quasar colors calculated for different filter combinations. The figure shows that the optimal redshift ranges is $\sim 6.0$ - 6.5 (up to 6.7). We did not push our quasar selection to a higher redshift to avoid a potential contamination from dwarf stars. When we selected quasars and calculated luminosity functions, we used the redshift ranges listed in Table 1.

In the second major step of our quasar selection, we performed a series of simulations on ACS images to demonstrate that we were able to select quasars with some extended far-UV radiation from host galaxies. The method has been introduced in the main text. The model galaxies that we generated have difference sizes (half-light radii = 0.1-0.3″; ref[56]) and Sérsic profiles (Sérsic index = 1 - 4). The dashed line in Fig. 2 shows one result from our simulations. In this case, the Sérsic index of the galaxy is 1, the half-light radius is 0.25″, and the galaxy flux is 30% of the total quasar flux. We also used spectroscopically confirmed galaxies (triangles in Fig. 2) to replace model galaxies in the above simulations, and the results are very similar.

### Quasar selection procedure in individual fields

We selected quasars in the fields listed in Table 1. These fields have deep imaging data that allow efficient target selections. In particular, they have deep HST ACS F814W or F850LP images with excellent PSFs of ~0.1″. Table 1 lists the data and filters for target selection in the individual fields. We will provide more details below.

The two GOODS fields[22] are the deepest fields here. In the optical, we used the HST ACS images (GOODS High-Level Science Products Version 2) in four bands $BViz$ (F435W, F606W, F775W, and F850LP) downloaded from the HST archive. The pixel size of the images (including all HST optical images used in this paper) is 0.03″. We used SExtractor[57] to do photometry in the dual mode, and the detection band was F850LP. We measured aperture magnitudes in a 0.27″ aperture in diameter. Aperture corrections were estimated using bright point sources, and were then applied to aperture magnitudes to obtain total magnitudes. In this paper, we use total magnitudes for all calculations. Strictly speaking, this total magnitude is not optimal for extended sources. It is adopted here for two main reasons. First, it is robust for point sources, including quasars (our targets). Second, it is reliable for relatively faint sources, because it is insensitive to the shape measurement of sources. Our selection limit is $z = 27.0$ mag, corresponding to a 10 $\sigma$ significance. This is the limit for a reliable FWHM measurement. We then selected $i$-band dropout ($i-z > 2$ mag) objects using the F775W−F850LP colors. We further required that these $i$-band dropouts should not be detected (< 3 $\sigma$ significance) in two bluer bands F435W and F606W. We visually inspected these objects and removed spurious detections. The final $i$-band dropouts in the two fields are shown as the blue circles in Fig. 2. Some $i$-band dropouts are spectroscopically confirmed galaxies at $z > 5.9$. Finally, we obtained $J$-band photometry of the $i$-band dropouts using the HST WFC3 F125W images produced by the CANDELS team[26]. No quasars were found in the GOODS fields down to our limit.

Inside the GOODS-S field, there are a few small regions (GOODS-S Deep in Table 1) that have exceptionally deep imaging data. We downloaded the data from the HST archive (https://archive.stsci.edu/prepds/hlf/) and selected quasars using the same procedure above. The only difference is that the magnitude limit is $z = 27.4$ mag. No quasars were found. Despite the great depth, the total area of these regions is only ~23 arcmin². When we calculated the area of the GOODS fields, we have excluded the area of these ultra-deep regions. In summary, we did not find quasars in the GOODS fields.

The COSMOS field[23] covers nearly two square degrees and is the largest imaging field in Table 1. This field is one of the two Subaru HSC ultra-deep survey fields[24], and has deep images in the $grizy$ bands. The field is also covered by HST ACS F814W-band images and by UltraVISTA near-IR images[25]. We used the HSC and UltraVISTA data for the color selection and used the F814W-band images to measure FWHMs. The detection band is the HSC $z$ band. We first used the HSC archive to search for point-like $i$-band dropouts. The resultant objects from this simple catalog search are mostly spurious detections, since real $i$-band dropouts are very rare. We then performed a forced photometry for these $i$-band dropout candidates on the HSC $gri$ images, and carried out the second selection of $i$-band dropouts. In this step, we required $i-z > 2$ mag and non-detection in $g$ and $r$ (the same criteria as we used for the GOODS fields). The magnitude limit is $z = 25.4$ mag. This is not limited by the depth of the $z$-band data (the HSC $z$-band images are much deeper). Instead, it is limited by the depth of the F814W-band images that were used to measure FWHMs. The 10 $\sigma$ depth of the F814W-band images is ~26.7 mag. For quasars at redshift between 6.0 and 6.5, their F814W−$z$(HSC) colors are in a very narrow range around 1.3-1.4 mag. Therefore, the $z$-band limit was set to be 25.4 mag, and the F814W−$z$(HSC) color was also used to remove contaminants. After we obtained the sample of $i$-band dropouts, we calculated their $J$-band photometry using the UltraVISTA images. No quasars were found in this field.

The depth of the F814W-band images in COSMOS is not uniform, and a fraction of them are shallow (no. 2c in Table 1). The depth of our quasar search in these shallow regions is $z \approx 25.0$ mag. The quasar selection procedure is the same as we did for the COSMOS field above. No quasars were found in these shallow regions.

The COSMOS CANDELS field is inside the COSMOS field. Although it is much smaller, it has much deeper F814W-band images from the CANDELS team. Therefore, the depth in the detection band was increased to $z \approx 26.1$ mag. When we measured aperture magnitudes on F814W-band images in this paper, we used a $0.24''$ aperture in diameter, which is slightly larger than the aperture size for point sources used by the CANDELS team[26], so the magnitude limits ($10\,\sigma$) quoted in Table 1 are slightly shallower than those provided by the CANDELS team[26]. The quasar selection procedure in this field is the same as we did for the COSMOS field. No quasars were found. When we calculated the area of the COSMOS field, we have excluded the shallow and CANDELS regions.

The EGS field[27] is inside one of the four CFHTLS fields and has deep $ugriz$ images. The whole EGS field is covered by HST F606W and F814W images, and the field can be divided into two parts based on the depth of the F814W images. The deep part (no. 3b in Table 1) is the EGS CANDELS field that has deep CANDELS images like other CANDELS fields. Our quasar selection was slightly different for the two parts. For the shallow part (no. 3a in Table 1), we used the CFHTLS data to select $i$-band dropouts and used the F814W-band images to identify point sources. We found no point-like $i$-band dropouts down to the magnitude limit $z = 25.4$ mag, so we did not use any near-IR images. For the deep part, the CFHTLS $z$-band image is not deep enough to match the F814W-band data, but its $i$-band image is deep enough. Therefore, we used $i$(CFHTLS) – F814W > 0.8 mag to select $i$-band dropouts and the detection band is F814W. We also required that $i$-band dropouts should not be detected in F606W. The HST WFC3 F125W image was used to identify L/T dwarfs. The magnitude limit is F814W = 27.5 mag, roughly corresponding to $z = 26.2$ mag. No quasars were found in this field.

The UDS CANDELS field is inside the UDS or SXDS field. The SXDS field is the other Subaru HSC ultra-deep survey field, and thus has deep images in the $grizy$ bands. In addition to the CANDELS images, this field is covered by the UKIDSS ultra-deep near-IR images[28]. The selection procedure is the same as we did for the COSMOS CANDELS field. No quasars were found in this field.

The HST GEMS field[29] is covered by ACS $V$ and $z$ bands (F606W and F850LP) over ~800 arcmin². The images were downloaded from the HST archive. The GOODS-S field is in the center of the GEMS field. In Table 1, the GEMS field (no. 5) is the region excluding the GOODS-S field. The GEMS field is also covered by Subaru HSC optical images and VLT VIDEO near-IR survey[30]. We downloaded HSC raw images from the archive and reduced them using hscPipe (https://hsc.mtk.nao.ac.jp/pipedoc/pipedoc_8_e/index.html). We found that only $i$-band images are deep enough for our purpose. The combination of the ACS, HSC, and VIDEO images were used for the quasar color selection. We first selected $V$-band dropout objects using the ACS $V$- and $z$-band images. We then performed forced photometry of the $V$-band dropouts on the HSC $i$-band images to select $i$-band dropouts. We found that the majority of the $i$-band dropouts in this step were cosmic rays (or spurious detections). In order to remove spurious detections, we combined VIDEO $J$, $H$, and $K$-band images to make a very deep, stacked, near-IR image. This image is deep enough to detect $z\sim6$ quasars down to our magnitude limit in this field. Finally, the color selection was done on the HSC $i$, ACS $z$, and VIDEO $J$ magnitudes. The ACS $z$-band images were used for the morphological selection. No quasars were found in this field.

Our survey area was mostly limited by deep ACS F850LP and F814W images available, because we required at least $10\sigma$ detections in these bands for reliable morphological measurements. We searched the HST archive for deep ACS F850LP or F814W images in the EGS, UDS, and COSMOS fields that have not been used by the fields mentioned above (no. 6 in Table 1). The result is listed in Extended Data Table 1. These data provided us with extra area. They were mostly from a few supernova surveys (HST program IDs are in the last column of the table). The quasar selection procedure is the same as we used earlier, and no quasars were found in these data. Note that we have excluded the area of these data in the area column in Table 1.

We further used our spectroscopic observations (no. 7 and 8 in Table 1) to search for faint quasars. In the first program[41], we observed LBGs over 340 arcmin$^2$ in the SDF field using Keck DEIMOS. The targets were $i$-band dropouts with $i - z > 1.7$ mag, and the integration time was ~3 hours per target. In the second program[40], we carried out a survey of high-redshift LAEs and LBGs over nearly two square degrees using Magellan M2FS. The LBG targets were $i$-band dropouts with $i - z > 1.5$ mag, and the integration time was 5-7 hours per target. We identified two samples of LAEs[58,59] but did not find any high-redshift quasars. The depth and completeness of the observations vary from field to field, but they are highly complete to $z \approx 25$ mag for type 1 quasars. First of all, our imaging data are much deeper than $z \approx 25$ mag. Second, bright targets with $z \leq 25$ mag had the highest priority, so nearly all of them were observed in our spectroscopic observations. Third, our spectroscopy was deep enough to identify quasars with $z \leq 25$ mag. Note that many of our spectroscopically confirmed LAEs have $z \sim 27$ mag[58,59]. Also note that these spectroscopic observations covered some of the fields in Table 1 and included many $i$-band dropouts in these fields.

**Calculation of the QLF and quasar contribution**

The basic method and the calculation steps are provided in the main text. To improve the constraint on quasar densities, we divide all fields into four groups based on their survey depths. The four groups are [1a, 1b, 6a, 6b], [2b, 3b, 4, 6c, 6d], [2a], and [2c, 3a, 7, 8], where the numbers correspond to the numbers in Table 1 and Extended Data Table 1. The survey volume of each group is the sum of the volumes of all individual fields in this group, and the effective $M_{1450}$ limit of this group is a volume-weighted average of individual $M_{1450}$ limits. The cumulative densities of bright quasars are computed from the differential densities in a previous study[16]. The results are shown in Fig. 3a as the triangles in steps of one magnitude except the faintest one with a step size of 0.5 magnitudes. When we perform a model fit to the fiducial QLFs, we add density errors to the upper limits of our density measurements. We assume that their relative errors are the median relative error of the quasar densities at the bright end. The area coverages of the two faintest data points are relatively small. To account for cosmic variance[60,61], we add extra errors of 30% and 15% in quadrature to these two points, respectively. The results are insensitive to how we add these errors.

Extended Data Table 1

| No | Field | Magnitude limit (mag) | Area (arcmin$^2$) | HST program IDs |
|----|-------|----------------------|-------------------|-----------------|
| 6a | UDS | $z \approx 26.8$ | 56 (6 pointings) | 9075, 12099 |
| 6b | COSMOS | $z \approx 26.8$ | 17 (2 close pointings) | 12461 |
| 6c | COSMOS | $z \approx 26.2$ | 62 (7 pointings) | 12461, 13294, 13641 |
| 6d | EGS | $z \approx 26.2$ | 102 (12 pointings) | 12547 |

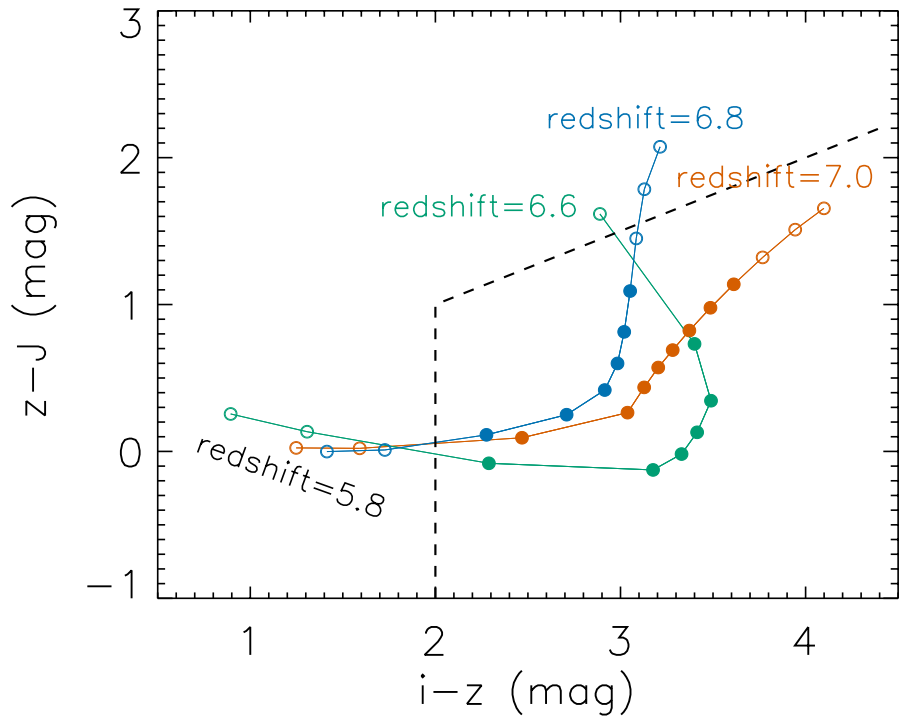

**Extended Data Fig. 1 The $z-J$ versus $i-z$ color-color diagram of high-redshift quasars for different filter sets.** The red, green, and blue circles show the median tracks of the quasar colors calculated for the GOODS, COSMOS, and EGS fields, respectively. The starting redshift is 5.8 and the step size is 0.1. The ending redshifts are different for different fields. The filled circles represent the redshift ranges that we used for target selection.

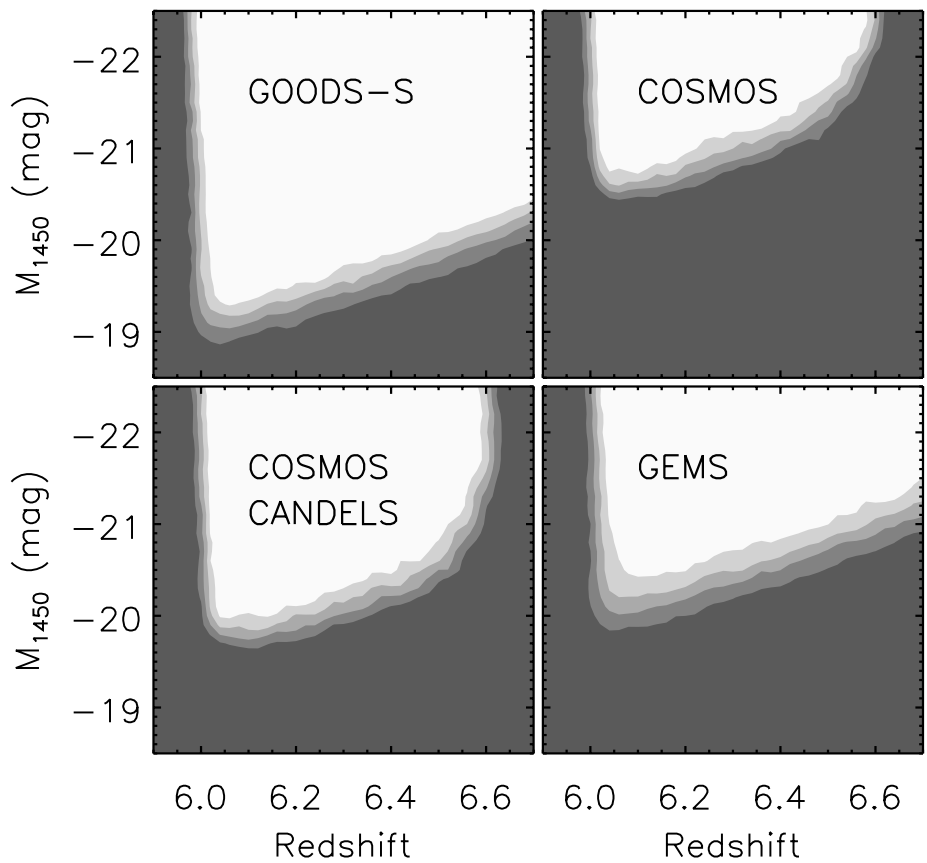

**Extended Data Fig. 2 Color selection completeness maps.** The contours are selection probabilities from 0.8 to 0.2 with an interval of 0.2 for four fields.